\documentclass[11pt,fleqn]{article}
\usepackage{amsmath,amssymb,graphicx}
\pdfoutput=1 
\usepackage{colortbl,xcolor}
\usepackage{algorithm}
\usepackage{algpseudocode}
\usepackage{verbatim}
\usepackage{todonotes} 
\begin{document}
\sloppy
\newtheorem{axiom}{Axiom}[section]
\newtheorem{claim}[axiom]{Claim}
\newtheorem{conjecture}[axiom]{Conjecture}
\newtheorem{corollary}[axiom]{Corollary}
\newtheorem{definition}[axiom]{Definition}
\newtheorem{example}[axiom]{Example}
\newtheorem{fact}[axiom]{Fact}
\newtheorem{lemma}[axiom]{Lemma}
\newtheorem{observation}[axiom]{Observation}
\newtheorem{proposition}[axiom]{Proposition}
\newtheorem{theorem}[axiom]{Theorem}
\newtheorem{defi}[axiom]{Definition}
\renewcommand{\topfraction}{1.0}
\renewcommand{\bottomfraction}{1.0}

\newcommand{\proof}{\emph{Proof.}\ \ }
\newcommand{\qed}{~~$\Box$}
\newcommand{\rz}{{\mathbb{R}}}
\newcommand{\nz}{{\mathbb{N}}}
\newcommand{\zz}{{\mathbb{Z}}}
\newcommand{\eps}{\varepsilon}
\newcommand{\cei}[1]{\lceil #1\rceil}
\newcommand{\flo}[1]{\left\lfloor #1\right\rfloor}
\newcommand{\seq}[1]{\langle #1\rangle}
\newcommand{\p}{\star}
\newcommand{\mf}[1]{\small{\emph{#1}}}
\newcommand{\mfb}[1]{\small{\color{blue}\bf{\emph{#1}}}}

\newcommand{\boxxx}[1]
 {\fbox{\begin{minipage}{11.80cm}\begin{center}\bigskip\begin{minipage}{11.30cm}
  #1\end{minipage}\end{center}~\end{minipage}}}
\newcommand{\ren}[1]{{\color{magenta}{#1}}}
       
\title{Recognising permuted Demidenko matrices} 
\author{
Eranda \c{C}ela\thanks{{\tt cela@opt.math.tu-graz.ac.at}.
Department of  Discrete Mathematics, TU Graz, Steyrergasse 30, A-8010 Graz, Austria}
\\Vladimir  Deineko\thanks{{\tt Vladimir.Deineko@wbs.ac.uk}.
Warwick Business School, The University of Warwick, Coventry CV4 7AL, United Kingdom}
\\Gerhard J.\ Woeginger\thanks{Deceased in April 2022.
 Lehrstuhl f\"{u}r Informatik 1 (Algorithmen und Komplexit\"{a}t), RWTH Aachen
 University, 
 Erweiterungsbau 1 (2353),  Ahornstr.\ 55, Germany}
}
\date{}
\maketitle
       
\begin{abstract}      

  We solve the recognition problem (RP)  for the class of Demidenko matrices.
  Our result closes a remarkable
  gap in the recognition of  specially structured matrices.
Indeed, the
recognition of permuted Demidenko matrices is a longstanding open problem, in
contrast to  the efficiently solved 
RP for important subclasses of Demidenko matrices such as  the  Kalmanson
matrices, the Supnick matrices, the Monge matrices and the Anti-Robinson
matrices.
   The recognition of the permuted Demidenko matrices is relevant in the context
of  hard combinatorial optimization problems   which become tractable if 
the input is  a Demidenko matrix.  
Demidenko matrices were introduced by  Demidenko  in 1976, when he proved that
the  Travelling  Salesman Problem (TSP)
is polynomially solvable if the symmetric distance       
matrix       
fulfills certain combinatorial conditions,        
nowadays known as  the {\em Demidenko conditions\/}.        
In the context of the TSP the recognition problem consists in deciding
whether there is a renumbering of  the cities such that the correspondingly renumbered 
distance matrix fulfills the Demidenko conditions, thus resulting in a
polynomially solvable special case of the TSP.        
We  show that such a renumbering of $n$ cities can be found  in $O(n^4)$  time, if it exists.
       
\medskip\noindent{\bf Keywords.}        
Combinatorial optimization, travelling salesman problem, Demidenko condition,       
permuted Demidenko  matrices.        
\end{abstract}       
       
\section{Introduction}\label{sec1}\nopagebreak       
Optimizing over permutations is a generic problem in combinatorial
optimization. In an instance  of size $n$,  the
set of feasible solution ${\cal F}_n$  is  a subset  of the set ${\cal S}_n$
of  permutations of $\{1,2,\ldots,n\}$, ${\cal F}_n\subseteq  {\cal S}_n$. 
The  generic problem $P$ is then given as   $\min \{f(\pi)\colon \pi \in {\cal
  F}_n\}$, where 
  $f\colon {\cal S}_n\to \rz$ is   the objective function.
Some fundamental NP-hard  problems in combinatorial optimization can be cast is
this way, as for example
{\em the travelling salesman problem (TSP)}, {\em the quadratic assignment
  problem (QAP)} or {\em the path travelling salesman problem  (PTSP)}.   
In all these cases the objective  function $f(\pi)$ is determined in terms of one or two matrices of coefficients.
In the TSP we are given as     input  an $n\times n$ distance matrix
$C=(c_{ij})$ and the objective  function $f^C_{TSP}$ is given as
\[f^C_{TSP}(\pi):=\sum_{i=1}^{n-1}c_{\pi(i)\pi(i+1)}+c_{\pi(n)\pi(1)}\, ,
  \mbox{ for }
\pi\in {\cal F}_n:={\cal S}_n\, .\]        
 In the QAP the input consists of two $n\times n$ matrices $A=(a_{ij})$,
 $B=(b_{ij})$ and the objective function $f^{A,B}_{QAP}$   is given as
 \[ f^{A,B}_{QAP}(\pi):=\sum_{i=1}^{n}\sum_{i=1}^{n} a_{\pi(i)\pi(j)}b_{ij}\, ,
 \mbox{ for }  \pi\in {\cal F}_n:={\cal S}_n\, .\]
 Finally, in the PTSP the input consists of  an $n\times n$ distance matrix $D=(d_{ij})$,  a
 start index $i$ and an end index $j$,   $i,j\in \{1,2\ldots,n\}$,  and the
 objective  function $f^D_{PTSP}$ is given as
 \[ f^D_{PTSP}(\pi):=\sum_{i=1}^{n-1}c_{\pi(i)\pi(i+1)}\, ,
 \mbox{ for } \pi \in {\cal F}_n:=\{ \pi \in {\cal S}_n\colon
 \pi(1)=i\, , \pi(n)=j\}\, .\]
 All these problems are   hard, both from  the theoretical and from the
 practical point of view.
 In particular, the TSP is one of the best studied problems in combinatorial optimization and
in  operational research, not only because of its  numerous practical
applications,   but also due to  its special  role in developing and testing
new approaches in the above mentioned fields, see e.g.~\cite{ABCC, Gutin, TSP}. 
The TSP  is NP-hard to solve exactly (see for instance 
\cite{GaJo}),  and APX-hard to approximate (see for instance 
\cite{PaYa93}).  The   QAP is a classical and  well studied problem with numeruous and
relevant applications.
It is  NP-hard to solve exactly   and NP-hard to approximate, see
eg.~\cite{SaGo76},  while  also being
very  hard from the practical point of view.
Indeed, solving   to optimality general  instances of  size $n=35$   still
remains   challenging, see for example the recent paper  \cite{Fujii_etal21}.  
Finally, the PTSP shares the (theoretical) hardness of the previously mentioned
problems being NP-hard and APX-hard, see eg.\ ~\cite{TVZ20,Zenk19}. 

Given the intractability of these problems, the characterization of tractable special cases is of 
obvious interest and forms a well-established and vivid branch of research.
Most of  the  tractable special cases  arise if certain combinatorial
conditions, as for example 
{\em four-point conditions} (see~\cite{DKTW}),  are imposed on the coefficient matrices of the problem.
These conditions give rise to special classes of matrices such as  for example  {\sl Monge}, {\sl Kalmanson},
{\sl  Supnick}, {\sl Anti-Robinson} and {\sl Demidenko}  matrices. There are
quite a number of  tractable special cases of the TSP  related to these matrix
classes,  we refer to 
\cite{DKTW} and the references therein for a comprehensive survey.
More recently,  four point conditions based special cases have also been
investigated  for   the QAP and the
PTSP, see  for example  \cite{Cela2018,CDW_TSP} and the references therein. 

Assume now that  the generic problem $P$ above is tractable (polynomially
solvable) if its coeffient matrix (or matrices) belongs to some
particular class (or classes) of matrices, say ${\cal C}$ (or ${\cal C}$ and
${\cal D}$).

Further,  consider an instance $I$  of $P$
with the following property: the  coefficient matrix (coefficient matrices) of
$I$ can be permuted according to some permutation $\varphi$ (or permutations
$\varphi$, $\psi$) so as to lie in ${\cal C}$. We refer to the instance of $P$
with the permuted coefficient matrix (coefficient matrices) as {\sl the permuted
instance}.  Notice that the permuted instance is tractable if the permutation
$\varphi$  (the permutations $\varphi$, $\psi$) is (are) known or can be efficiently computed. In
this case     an optimal
solution of the original instance $I$ can be easily obtained
from the optimal solution of the permuted  instance by using   the permutation $\varphi$
($\varphi$ and $\psi$).  Let us illustrate the idea in terms of the TSP.
Assume that the  special case of the TSP where the   distance matric belongs to
${\cal C}$ is polynomially solvable.  Consider an instance of the
TSP with  distance matrix $C$ which 
can be permuted, say  by a permutation $\varphi$, such that
$C^{\varphi}:=(c_{\varphi(i)\varphi(j)})\in {\cal C}$. Let $\pi^{\ast}$ be
  an  optimal solution of  the TSP instance with
  distance matrix $C^{\varphi}$. Then 
  $\varphi\circ\pi^{\ast}$ is an optimal solution of the TSP instance with
  distance matrix $C$ because the following equality hods for any $\pi \in {\cal
    S}_n$:
  \[ f_{TSP}^{C^{\varphi}}(\pi)=\sum_{i=1}^{n-1}c^{\varphi}_{\pi(i)\pi(i+1)}+c^{\varphi}_{\pi(n)\pi(1)}=
    \sum_{i=1}^{n-1}c_{\varphi(\pi(i))\varphi(\pi(i+1))}+c_{\varphi(\pi(n))\varphi(\pi(1))}=f_{TSP}^C(\varphi\circ\pi)\, \]

Thus,  the question     whether there exists a permutation $\varphi$ such that
$C^{\varphi}\in {\cal C}$ holds is relevant for the efficient solvability of
the considered TSP instance. Being able to answer this question and to determine the corresponding permutation
in the positive case would result in a larger class of polyniomially
solvable special cases of the problem. 

The generic recognition problem for a class ${\cal C}$ of matrices (which could
be described in terms of some combinatorial properties) is defined as follows:
\begin{center}
\boxxx{\textbf{Problem:}  {\sc Recognition}-${\cal C}$
\\[1.0ex]
\textbf{Instance:} $n\in \nz$, an $n\times n$ matric $C$.
\\[1.0ex]
\textbf{Task:} Is there some permutation $\varphi \in {\cal S}_n$ such that
the permuted matrix $C^{\varphi}$ belongs to ${\cal C}$? If yes, then determine
such a $\varphi$. }
\end{center}
\smallskip

In particular, the recognition problem for the class of Demidenko matrices is relevant in
the context of the following two results. In 1976,  Vitali Demidenko~\cite{Demi}       
proved that the TSP restricted to Demidenko matrices can be solved in       
$O(n^2)$ time (cf.\ Gilmore, Lawler and Shmoys~\cite{GLS}).  
In the recent paper \cite{CDW_TSP} it was shown that the PTSP on 
Demidenko matrices is also solvable  in polynomial time. 
\smallskip

While the recognition problem for the class of Demidenko matrices has been open
for around fourty years, it is known to be polynomially solvable for a number of proper
sublasses of the class of Demidenko matrices such as the  Kalmanson
matrices, the Supnick matrices, the Monge matrices and the Anti-Robinson
matrices, see \cite{DF,DRW2,DRW,LS,PF}. All these classes of matrices give rise
to polynomially tractable cases of combinatorial optimization problems over
permutations,  including  the  TSP, the QAP and/or the PTSP, see for example \cite{BSurv,BKR,Cela2018,CDW_TSP,DKTW,DRW,GLS,Kabadi,K,LS15,PSW,S}. 
In this paper we close the gap and 
solve the recognition problem for Demidenko matrices. More precisely, we show how to decide
whether   a given  $n\times n $ matrix $D$ can be permuted to a Demidenko
matrix and how to construct the corresponding permutation in $O(n^4)$ time,  in the positive
case. 

\smallskip
\noindent{\bf Organization of the paper.}       
 In the next section we introduce some basic notations and   define the
 Demidenko matrices as well as    the
 Anti-Robinson matrices which turn out to be relevant in this paper. In
 Section~\ref{sec:relationship}
 we  discuss some properties of (permuted) Demidenko and (permuted) Robinson
 matrices and focus on the relationship between these matrix classes.
 Then, in Section~\ref{sec:algo},   we present an 
algorithm for  Recognition-${\cal D}$,  the recognition problem  for the class
${\cal D}$ of Demidenko matrices.      
Finally we close the paper with a short summary in Section~\ref{sec:concl}.  
\section{Definitions and notations}\label{sec:defi}\nopagebreak 
      
\begin{defi} A symmetric $n\times n$-matrix $C = (c_{ij})$ is called a
 {\em Demidenko matrix\/} if its entries satisfy the inequalities      
\begin{equation} \label{dem.s} 
c_{ji} + c_{kl} \leq c_{jl} + c_{ki} \, ,      
\mbox{\qquad for all $1\le i < j < k < l\le n$.}      
  \end{equation}
A       matrix $C=(c_{ij})$ is called a  {\em permuted}       
Demidenko  matrix if there is a permutation $\varphi$        
 of its rows and columns such that the permuted matrix       
 $C^{\varphi}=(c_{\varphi(i)\varphi(j)})$  is a Demidenko     matrix.
 Such a permutation $\varphi$ is called a {\em Demidenko    
   permutation for  the  matrix $C$\/}.
\end{defi}

Note that the conditions (\ref{dem.s}) do not  involve the         
entries $c_{ii}$,  $c_{1i}$ and
$c_{ni}$,  for $i \in \{1,2,3,\ldots,n\}$, respectively.        

The system (\ref{dem.s}) contains $O(n^4)$ inequalities, but it can be easily  seen
that it
is equivalent to the following  system  of  $O(n^3)$ inequalities 
   \begin{equation}\label{eq:demi}      
c_{ji} + c_{j+1,l} \leq c_{jl} + c_{j+1,i}   \, ,    
\mbox{\qquad for all $1\le i < j < j+1 < l\le n$.}      
 \end{equation}    
 Further,  it can be  easily seen, that the later system (\ref{eq:demi}) is
 equivalent to the following conditions which can be checked in $O(n^2)$ time 
 \begin{eqnarray}       
\max_{i=1,\ldots,j-1}\{c_{ji}-c_{j+1,i}\} \le       
\min_{l=j+2,\ldots,n}\{c_{jl}-c_{j+1,l}\}\, ,        
&& \mbox{~for all~~} 2\le j\le n-2.  
\label{dem.r}    
\end{eqnarray}       
Thus, it can be decided in $O(n^2)$ time  whether a given symmetric $n\times n$ matrix  $C$ is a
Demidenko matrix.       
\smallskip      

Next we define  the Robinson matrices. They arise in 
combinatorial data analysis~\cite{Robinson} and have some nice applications and
implications in the context of combinatorial  optimization problems, see e.g.\  \cite{Cela2018,LS15,PF}.
\begin{defi}\label{defi:Robi}
  A symmetric $n\times n$ matrix $A=(a_{ij})$ is called an  \emph{Anti-Robinson
    matrix} if  it satisfies the following conditions:
\begin{equation}
a_{ik}\ge \max \{a_{ij},a_{jk}\}\, ,   \mbox{\qquad for all  $1\le i<j<k\le n.$} \label{rob.c}
\end{equation}
In words, in each row and  column of $A$ the entries
do not  decrease when moving away from the main diagonal.
 A       matrix $C=(c_{ij})$ is called a  {\em permuted}       
Anti-Robinson   matrix if there is a permutation $\varphi$        
 of its rows and columns such that the permuted matrix       
 $C^{\varphi}=(c_{\varphi(i)\varphi(j)})$  is an Anti-Robinson matrix.       
Such a permutation $\varphi$ is called an {\em Anti-Robisnson     
  permutation for  the  matrix $C$\/}.      
\end{defi}

Notice that  the Anti-Robinson matrices build  a subclass
of the class of 
the Demidenko matrices.
Indeed, the inequalities~(\ref{rob.c}) imply $a_{ji}\le a_{ki}$ and $a_{kl}\le
a_{jl}$,  for  $1\le i<j<k<l\le n$, and  by summing up
these inequalities we  obtain the 
inequalities~(\ref{dem.s}).  It is easy to check that the  inclusion mentioned above   is proper,
i.e.\ that there are Demidenko matrices which are not Anti-Robinson matrices. 
See for example the  $5\times 5$ Demidenko  matrix  below and notice that
similar $n \times n$ matrices can be constructed for any $n\in \nz$.  
\[\left ( \begin{array}{ccccc}
            0&1&0&0&0\\
            1&0&0&1&1\\
 0&0&0&0&0\\
            0&1&0&0&0\\
               0&1&0&0&0\end{array}\right  )\, .\]
         Since $a_{13}<a_{12}$, $A$ is not an Anti-Robinson matrix.




\smallskip

\noindent{\bf Notations.}         
In the following  we will introduce some notations used throughout the paper. 
  For a given  $n\times n$ matrix $C$ we  identify each of its rows (columns) by
  the corresponding index. Thus        
$I=\{1,\ldots,n\}$ is the set of rows (columns) of $C$.  A row $i$       
{\em precedes} a row $j$ in $C$ ($i\prec j$,  for short),        
if row $i$ occurs before row $j$ in $C$.       
A set  $K_1$ of rows (columns) is said to precede a  set $K_2$ of rows
(columns)  iff $k_1\prec k_2$ for all $k_1\in K_1$       
and for all $k_2\in K_2$.  In this case  we write $K_1\prec       
K_2$. 
Let $V=\{v_1,v_2,\ldots,v_r\}$  be a subset of $I$. We    
denote by $C[V]$ the $r\times r$ submatrix of $C$ which   
is obtained by deleting all rows and columns not contained in $V$.           
      
\smallskip       
For a permutation $\pi \in {\cal S}_n$, we
denote 
$\pi=\seq{x_1,x_2,\ldots,x_n}$  iff $\pi(i)=x_i$ holds for    all    
$i\in \{1,2,\ldots,n\}$. 
%
For two subset of indices  $K,L\subset I$ with $K\cap
L=\emptyset$ and a permutation $\pi$ we  say that $K$ precedes $L$ in $\pi$
iff for any $k\in K$ and for any $l\in L$,  the indices
$i, j\in I$ with  $\pi(i)=k$, $\pi(j)=l$ fulfill $i<j$.
In this case we write  $K\prec_{\pi} L$ or simply  $K\prec  L$, 
whenever the permutation $\pi$  is  clear from the context.
Finally, let  $I=\{1,2,\ldots,n\}$ and $ K \subset I$
with $K=\{s_1,s_2,\ldots,s_{|K|}\}$ auch that  $1<|K|<n$ and $s_1<s_2<\ldots <s_{|K|}$.
We say that a permutation $\tau \in {\cal S}_n$ and a permutation $\sigma \in {\cal
  S}_{|K|}$ coincide on $K$ iff for any $i,j\in \{1,2,\ldots,|K|\}$, $\tau (s_i)<\tau(s_j)$ implies
$\sigma(i)<\sigma(j)$ and vice-versa.  

\section{The relationship between permuted  Demidenko matrices   and permuted
  Anti-Robinson matrices}     
\label{sec:relationship}\nopagebreak       
      
In this section we investigate the relationship between (permuted)  Demidenko
matrices and (permuted) Anti-Robisnon matrices. 
The next lemma describes the    relationship between some special Demidenko
matrices and Anti-Robinson matrices.
\begin{lemma}\label{DemiRobi:lem1}
  Let $C$ be an $n\times n$ Demidenko matrix such $c_{1j}=c_{i1}=a$ for some
  $a\in \rz$ and any $i,j\in I:=\{2\ldots,n\}$.
  Further, assume that for some $r\in \{1,2,\ldots,n-2\}$ there exists constant
  $b\in \rz $ and a subset of indices $K\subseteq I':=I\setminus \{1,2,\ldots,r,n\}$ such
  that  $\sum_{j=1}^r c_{ij}-rc_{in}=b$ for any $i \in K$.
  Then the submatrix $C[K]$ is an Anti-Robinson matrix.
\end{lemma}
\proof Consider  first any two indices $i_i,i_2\in K$, $i_1 < i_2$ and observe that 
 \begin{equation}\label{equal:sum}
 \sum_{j=1}^r c_{i_1j}-rc_{i_1n}=\sum_{j=1}^r c_{i_2j}-rc_{i_2n}
  \end{equation}
 implies $c_{i_1j}-c_{i_1n}=c_{i_2j}-c_{i_2n}$ for any $j \in \{1,2,\ldots,r\}$.
 Indeed, assume that  at least one of these equalities does not hold and
 denote by $j_1$ the
 smallest index in $\{1,2,\ldots,r\}$ for which $ c_{i_1j_1}-c_{i_1n}\neq c_{i_2j_1}-c_{i_2n}$. 
The Demidenko condition for the quadruple $j_1<i_1<i_2<n$ implies $
c_{i_1j_1}-c_{i_1n}< c_{i_2j_1}-c_{i_2n}$.
The latter inequality together with \ref{equal:sum} implies the existence of a
$j_2\in \{1,2,\ldots,r\}$,
such that $ c_{i_1j_2}-c_{i_1n} >c_{i_2j_2}-c_{i_2n}$ holds.
But this is a violation of  the Demidenko condition for the quadruple $j_2<i_1<i_2<n$. 
In particular for $j=1$ denote  $c_{i_11}-c_{i_1n}=c_{i_21}-c_{i_2n}=:b'$ for
any $i_1,i_2\in K$.
The latter equalities   together with $c_{i_11}=c_{i_21}=a$ imply $c_{in}=c_{ni}=a-b'$ for any $i \in K$.
Now consider a triple   $j,k,l \in K$ with $j<k<l$ and apply
(\ref{dem.s}) with $i=1$:  $c_{j1}+c_{kl}\le c_{k1}+c_{jl}$.  With
$c_{j1}=c_{k1}$ we get $c_{kl}\le c_{jl}$.
Analogously, for any triple    $i,j,k \in K$ with $i<j<k$ we apply
(\ref{dem.s}) with $l=n$:  $c_{ji}+c_{kn}\le c_{jn}+c_{ki}$.  With
$c_{kn}=c_{jn}=a-b'$ we get $c_{ji}\le c_{ki}$.
Thus the submatrix $C[K]$ fulfills  the inequalities (\ref{rob.c}) in Definition~(\ref{defi:Robi}),
therefore  it is   an
Anti-Robinson matrix.
\qed

The relationship stated by the above lemma extends to a relationship
between permuted Demidenko and permuted Anti-Robinson matrices as follows.

\begin{lemma}\label{DemiRobi_b:lem}
  Let $C$ be an $n\times n$  permuted Demidenko matrix and let $\tau\in {\cal
    S}_n$ be a   Demidenko permutation for $C$. Assume that $\tau$  fulfills
  $\tau(n)=q$  and $\tau(j)=p_j$ for $j\in \{1,2,\ldots,r\}$ and some  $r\le
  n-2$, where   $p_1,p_2,\ldots,p_r,q\in I:=\{1,2,\ldots,n\}$ are pairwise different indices.
  Moreover, let $C$ have a constant first row (column), i.e.  $c_{1i}=c_{i1}=a$ hold for some
  $a\in \rz$   and any $i\in \{2,\ldots,n\}$.
   Further, assume that there exist a constant
  $b\in \rz $ and a subset of indices $K\subseteq I':=I\setminus \{p_1,\ldots,p_r,q\}$ such
  that  $\sum_{j=1}^r c_{\tau(i)p_j}-rc_{\tau(i)q}=b$ for any $i \in K$.
  Then, the submatrix $C[K]$ is a permuted  Anti-Robinson matrix. 
   In particular,  the unique  permutation $\sigma \in {\cal S}_{|K|}$
  which coincides with $\tau$   on $K$ is an Anti-Robinson permutation   for  $C[K]$.
\end{lemma}
\proof
Consider the Demidenko matrix $C^{\tau}=(c^{\tau}_{ij})$ with
$c^{\tau}_{ij}=c_{\tau(i)\tau(j)}$ for any $i,j \in I$. Clearly, all
non-diagonal entries
in its first row and column are equal to $a$.  Further, for any
$i \in K$ we have $\sum_{j=1}^rc^{\tau}_{ij}-rc^{\tau}_{in}=\sum_{j=1}^rc_{\tau(i)p_j}-rc_{\tau(i)q}
=b$. Lemma~\ref{DemiRobi:lem1} implies that the matrix $C^{\tau}[K^{\tau}]$ is an
Anti-Robinson matrix.
Now observe that $C^{\tau}[K^{\tau}]$ results from $C[K]$
by permuting its rows and columns according to the unique permutation $\sigma \in {\cal S}_{|K|}$
which coincides with $\tau$ on $K$.
\qed 

Next observe that we can   simply transform a given Demidenko matrix into a
Demidenko matrix with a constant first row (and first column).
\begin{observation}\label{obse}
  Let $C$ be an $n\times n$  Demidenko matrix. Then the matrix $C^{\prime}=(c_{ij}^{\prime})$
  with $c^{\prime}_{ij}:=c_{ij}-c_{1j}-c_{i1}$ for any  $i,j\in I:=\{1,2,\ldots,n\}$ is a
  Demidenko matrix with $c^{\prime}_{i,1}=c^{\prime}_{1,i}=-c_{11}$ for any
  $i \in I$. 
\end{observation}
Let $C$ be  a   given  {\em  symmetric\/} $n\times n$  matrix. 
The goal is to  decide  whether there  exists a Demidenko permutation      
$\tau$ for the matrix $C$, i.e.\  a permutation $\tau$ such that
$C^{\tau}=(c_{\tau(i)\tau(j)})$ is a Demidenko matrix,
and  to compute $\tau$, if  it exists.
To this end, we will  identify   some simple
combinatorial properties of  Demidenko permutations for the matrix
$C$.
\begin{lemma}\label{lemm:firstorder1}
  Let $C$ be  a     {\em  symmetric\/} $n\times n$  matrix and let $p_1,p_2,\ldots,p_r,q\in I:= \{1,2,\ldots,n\}$,
  be pairwise different  indices for  $r \le n-2$. Let $I'=I\setminus
  \{p_1,\ldots,p_r,q\}$ and
  $m:=\min \{ \sum_{j=1}^r c_{ip_j}-rc_{iq}\colon i
  \in I' \}$. 
   Further let  $K:=\{ i \in
  I'\colon \sum_{j=1}^k c_{ip_j}-kc_{iq}=m\}$.
For any   Demidenko permutation  $\tau$ for  $C$   such that
$\tau(j)=p_j$, for $1\le j\le r$,  and
  $\tau(n)=q$  (if such a permutation exists) the following  statements hold:  
  \begin{itemize}
    \item[(i)] If  $K=\{s\}$ for some $s\in I'$,
      then $\tau(r+1)=s$.
      
    \item[(ii)]  If $1<|K|<n-r-1$, then $K\prec L$ in $\tau$,  where
      $L:=I'\setminus   K$. In other words,
      $\{\tau(i)\colon i\in \{r+1,\ldots,r+|K|\}\}=K$ holds.
    \end{itemize}
  \end{lemma}
\proof 
Since $C^{\tau}$ is a Demidenko matrix 
  with $\tau(j)=p_j$, for $1\le  j\le r$, and   $\tau(n)=q$, we  apply
  (\ref{dem.s}) with $i\in \{1,2\ldots,r\}$, $j,k\in I'$ such that $j<k$, $l=n$
  and
  obtain 
  \[ c^{\tau}_{ji}-c^{\tau}_{jn}\le c^{\tau}_{ki}-c^{\tau}_{k n}\mbox{ or
      equivalently } c_{\tau(j)p_x}-c_{\tau(j)q}\le c_{\tau(k)
      p_x}-c_{\tau(k) q} \, .   \]
  This inequalities imply
  \begin{equation}\label{perm:Demi}
    \sum_{x=1}^r c_{\tau(j),p_x}-rc_{\tau(j),q}\le \sum_{x=1}^r c_{\tau(k),p_x}-rc_{\tau(k),q}
    \mbox{ for
      any $k>j$, $k,j\in I'$}\, .  \end{equation}
  The inequalities  (\ref{perm:Demi}) imply $\tau(r+1)\in K$.
  Then, for   $K=\{s\}$ we get $\tau(r+1)=s$ and thus (i) holds. 

   Assume now  $1<|K|\le n-2$. The inequalities  (\ref{perm:Demi}) implies that $\{s\}$
  precedes $\{s'\}$ in $\tau$  for  any $s\in
  K$ and any $s'\in L=I'\setminus K$. Thus $K\preceq L$ in $\tau$
  and (ii) holds.
\qed

\begin{lemma}\label{lemm:Robinsoncharact1}
Let $C$ be  a     {\em  symmetric\/} $n\times n$  matrix and let $p_1,p_2,\dots,p_r,q\in I:= \{1,2,\ldots,n\}$,
  be pairwise different indices for some $r$ fulfilling  $1\le r\le n-3$.  
  Let $m:=\min \{ \sum_{x=1}^r c_{ip_x}-rc_{iq}\colon i
  \in I':=I\setminus \{p_1,\ldots,p_r,q\}\}$,  
    $K:=\{ i \in
   I'\colon \sum_{x=1}^r c_{ip_x}-rc_{iq}=m\}$ and  $L:=I'\setminus K$.
   Let $C'=(c'_{ij})$ be defined by  $c'_{ij}=c_{ij}-c_{ip_1}-c_{p_1j}$, for $i,j
   \in I$.
   Assume further the   existence of a Demidenko permutation $\tau$ with respect
        to $C$  such that $\tau(j)=p_j$ for $j\in \{1,2,\ldots,r\}$  and
        $\tau(n)=q$.
\begin{itemize}
    \item[(i)] If $1<|K|=n-(r+1)$, then  $C'[K]$ is a permuted Anti-Robinson
      matrix.
      Moreover, for any Anti-Robinson permutation $\pi$ with
        respect to $C'[K]$,  the  permutation $\tau'\in {\cal S}_n$  which
        coincides with $\pi$ on $K$ and 
         fulfills $\tau'(x)=p_x$ for $x\in \{1,2,\ldots,r\}$ and  $\tau'(n)=q$, is a Demidenko permutation
         for $C$. 
   %
    \item[(ii)] Assume   $1<|K|<n-(r+1)$.  Consider the symmetric  $(|K|+1)\times (|K|+1)$
            matrix $D=(d_{ij})$ obtained  from  $C'[K]$ by appendig to
           it a $(|K|+1)$-st  column  (row) with arbitrarily chosen
           $d_{|K|+1,|K|+1}$ and    $d_{i,|K|+1}=d_{|K|+1,i}=M\sum_{j\in
             L}c_{ij}$ for   $i\in \{1,2,\ldots,|K|\}$ and  $M$ being a
                                positive large  constant.
            Then, $D$ is   a permuted
           Anti-Robinson matrix and an Anti-Robinson permutation
           $\sigma'$ for matrix $D$ is obtained by  $\sigma'(|K|+1)=|K|+1$ and
           $\sigma'(i)=\sigma(i)$ for any $i\in \{1,2,\ldots,|K|\}$, where
           $\sigma$ is the unique permutation which coincides with $\tau$ on
           $K$.
           
           Vice-versa, for any  Anti-Robinson permutation $\pi'$  for matrix $D$,
           there exists a Demidenko permutation $\tau'$ for matrix  $C$ such
 that  $\tau'(x)=p_x$ for $x\in \{1,2,\ldots,r\}$,  $\tau'(n)=q$ and $\tau'$
 concides with $\pi'$ on $K$.  
   \end{itemize}
   \end{lemma}
 {\sl Proof of (i).}
 Assume first that  $|K|=n-(r+1)$.
 Then,   Lemma~\ref{DemiRobi_b:lem} implies that $C'[K]$ is a permuted
 Anti-Robinson matrix and that    the uniquely defined   permutation $\sigma\in {\cal S}_{|K|}$ which
        coincides with $\tau$ on $K$ is an   Anti-Robinson permutation for         matrix  $C'[K]$.

        Consider now any  Anti-Robinson permutation
        $\pi$ for matrix  $C'[K]$ and let $\tau'\in {\cal S}_n $ be the uniquely defined  permutation which
        coincides with $\pi$ on $K$ and 
         fulfills $\tau'(i)=p_i$, for $1\le i\le r$, and  $\tau'(n)=q$. It can be easily checked that
         $C'$ permuted by $\tau'$ fulfills the Demidenko conditions
         (\ref{eq:demi}). Indeed for any $i,j,l$ in $I'$ with  $r<i<j<j+1<l<n$
         we have
         \begin{equation}\label{equ:auxiliary}
         c'_{\tau'(j)\tau'(i)}\le c'_{\tau'(j+1)\tau'(i)} \mbox{ and }
         c'_{\tau'(j+1)\tau'(l)}\le c'_{\tau'(j)\tau'(i)}\, ,
         \end{equation} because $\tau'$ coincides
         with $\pi$ on $K$ and $C'[K]$ permuted by $\pi$ is an Anti-Robisnosn
         matrix. The  inequalities (\ref{equ:auxiliary}) clearly imply  (\ref{eq:demi}) in this
         case.
         Further, for any $i\le r$, and for any $j,l\in I'$ with  $i<j<j+1<l<n$ 
        the leftmost  inequality in (\ref{equ:auxiliary}) is fulfilled by equality,
        whereas for  any $i,j\in I'$ with  $r<i<j<j+1<l=n$ the rightmost
        inequality in  (\ref{equ:auxiliary}) is fulfilled by equality. Finally
        for any $i\le r$ and for any $j\in I'$ with  $i<j<j+1<l=n$ both inequalities  in
        (\ref{equ:auxiliary}) are  fulfilled by equality. Thus $C'$
        permuted by $\tau'$ fulfills the Demidenko conditions for any $i,j,l$ in
        $I$ with  $1\le i<j<j+1<l\le n$ and this completes the proof of (i). 
        \smallskip
        
        {\sl Proof of (ii).}
    Consider now the case  $1<|K|<n-(r+1)$. Let   $K=\{s_1,s_2,\ldots,s_{|K|}\}$ and
    $s_1<s_2<\ldots < s_{|K|}$. Consider the  Demidenko permutation
    $\tau$ for matrix   $C$ with $\tau(x)=p_x$ for $x\in \{1,2,\ldots,r\}$ and $\tau(n)=q$.
    Then,  obviously, $\tau$ is also a Demidenko permutation for  $C'$.
    Moreover $c'_{\tau(1)i}=c'_{i\tau(1)}=-c_{pp}$ for any
    $i \in I$. Further,  $\sum_{x=1}^rc'_{\tau{i}p_x}-rc'_{\tau(i)q}=m$ for any $i\in K$  due
    to the definition of $K$.
    Then, Lemma  \ref{DemiRobi_b:lem} implies
  that $C'[K]$ is a permuted Anti-Robinson matrix and the uniquely defined   $\sigma\in {\cal
    S}_{|K|}$ which coincides with $\tau$ on $K$ is an Anti-Robinson
  permutation for $C'[K]$. Consider now the  $(|K|+1)\times (|K|+1)$ matrix $D$.
  We show that it is 
  permuted Anti-Robinson matrix. More precisely, we show that 
  the permutation  $\sigma'\in {\cal S}_{|K|+1}$ given by $\sigma'(i)=\sigma (i)$ for
  any $i \in \{1,2,\ldots, |K|\}$ and $\sigma'(|K|+1)=|K|+1$ is an
  Anti-Robinson permutation for $D$.
To this end  it is enough to show the following two families of  inequalities
  \begin{equation}
    \label{ARob1} d_{\sigma'(i+1),|K|+1} \le
    d_{\sigma'(i),|K|+1} \mbox{ for any $i\in \{1,2,\ldots,|K|-1\}$}
  \end{equation}
  and
  \begin{equation}\label{ARob2} d_{\sigma'(i)\sigma'(|K|)}\le
    d_{\sigma'(i),|K|+1} \mbox{ for any $i\in \{1,2,\ldots,|K|-1\}$} \, .
    \end{equation}
Notice that for any $i\in \{1,2,\ldots,|K|-1\}$ we have
\begin{equation}
  \label{ARob1:aux1}
  d_{\sigma'(i+1),|K|+1}=d_{\sigma(i+1),|K|+1}=d_{\tau(s_{i+1}),|K|+1}=M\sum_{j\in
    L} c'_{\tau(s_{i+1})\tau(j)}\, ,
\end{equation}
and analogously
\begin{equation}\label{ARob1:aux2}
  d_{\sigma'(i),|K|+1}=M\sum_{j\in
    L} c'_{\tau(s_{i})\tau(j)} \, .
\end{equation}
Since $K\preceq L$ and the matrix $(C')^{\tau}$ (obtained through the
permutation of  $C'$  by $\tau$)
is a Demidenko matrix with a constant first column,  the Demidenko conditions
 for  $(C')^{\tau}$ with $1<i<i+1<j$                imply
$c'_{\tau(s_{i+1})\tau(j)}\le c'_{\tau(s_{i})\tau(j)}$ for any $j\in
L$. The latter inequalities together with  (\ref{ARob1:aux1}) and
(\ref{ARob1:aux2}) imply (\ref{ARob1}). Finally,  (\ref{ARob2}) is equivalent to
$c'_{\tau(s_i)\tau(s_{|K|})}\le M\sum_{j\in L} c'_{\tau(s_i)\tau(j)} $ and the
latter inequality 
can be guaranteed by choosing $M$ large enough. 
\smallskip

Now assume that $D$ is an Anti-Robinson matrix and consider an Anti-Robinson
permutation $\pi'\in {\cal S}_{|K|+1}$ for matrix  $D$. By choosing $M$
large enough we can ensure that $\pi'(|K|+1)=|K|+1$. Then,  $\pi \in {\cal
  S}_{|K|}$ with $\pi(i)=\pi'(i)$ for any $i\in \{1,2,\ldots,|K|\}$ is an
Anti-Robinson permutation for $C'[K]$.
Consider the  Demidenko permutation  $\tau$ for $C$ with 
$\tau(j)=p_j$ for $j\in \{1,2,\ldots,r\}$ and $\tau(n)=q$ which exists
according to the assumptions of the lemma.
Clearly,  $\tau$ is also a Demidenko permutation for matrix  $C'$.
Hence, Lemma~\ref{lemm:firstorder1} implies that $K\preceq
L$ in $\tau$, i.e.\ $\{\tau(i)\colon i \in K\}=\{r+1,\ldots,,|K|+r\}$ holds. If
$\tau$ coincides with $\pi$ on $K$ we are done.
Otherwise, we  modify $\tau$ by permuting the entries $\tau(s_i)$, $i \in K$, so as  to
obtain  a new permutation $\tau'\in {\cal S}_n$  which coincides  with $\pi$ on
$K$ and  fulfills $\tau'(j)=p_j$,  for $j\in \{1,2,\ldots,r\}$, as well as  $\tau'(n)=q$.
The proof is completed by showing  that $\tau'$ is a Demidenko permutation for
$C'$ and hence also for $C$.

Assume w.l.o.g.\  that $\tau'$ is obtained by applying to $\tau$ a
transposition $(s_i,s_j)$ for some $i,j\in \{1,2,\ldots,|K|\}$, with $i\neq
j$.
This means that $\tau'(s_i)=\tau(s_j)$, $\tau'(s_j)=\tau(s_i)$ and
$\tau(l)=\tau'(l)$ for any $l\in I\setminus\{s_i,s_j\}$.
We  show that  the rows (columns) $s_i$ and $s_j$ of
$C'$ coincide except for maybe the  diagonal
entries; this coincidence   would then  imply that $\tau'$ is a Demidenko permutation
for $C'$ and also for $C$.

Assume by contradiction that there exist an $l \in
I\setminus \{s_i,s_j\}$ such that $c'_{s_i,\l}\neq c'_{s_j,l}$. Let
$\sigma'\in {\cal S}_{|K|+1}$ and $\sigma\in {\cal S}_{|K|}$ be  defined from
$\tau$  as  in 
the
already proved direction of (ii). Thus $\tau$ coincides with $\sigma$ on
$K$,  $\sigma$ is an Anti-Robinson permutation for matrix $C'[K]$ and
$\sigma'$ is an Anti-Robinson permutation for matrix $D$. 
Since  $\tau'$ is obtained  by applying to $\tau$ a
transposition $(s_i,s_j)$, then $\sigma$  is obtained from $\pi$ and $\sigma'$
is obtained from $\pi'$ by applying the transposition $(i,j)$, respectively.
Observe that $l \not\in K$. Indeed, under the assumption   $l\in K$ and
by  assuming w.l.o.g.\ $\tau(s_i)<\tau(s_j)<\tau(l)$
 we get 
$c'_{\tau(s_j)\tau(l)}\le  c'_{\tau(s_i)\tau(l)}$ and $c'_{\tau(s_i)\tau(l)}=c'_{\tau'(s_j)\tau(l)}\le
c'_{\tau'(s_i)\tau(l)}=c_{\tau(s_i)\tau(l)}$, hence 
$c'_{\tau(s_i)\tau(l)}=c'_{\tau(s_j)\tau(l)}$, a contradiction to
$c'_{s_i,l}\neq c'_{s_j,l}$.
Thus $l\not \in K$,  implying $l \in L$. Since $\pi'$
 and $\sigma'$ are both Anti-Robinson permutations for $D$ and
 $\pi'(k)=\sigma'(k)$ for any $k \not\in \{i,j\}$ we get by an analogous
 argument that $d_{i,|K|+1}=d_{j,|K|+1}$,  or equivalently,  $\sum_{t\in
   L}c'_{\tau(s_i)\tau(t)}=\sum_{t\in
   L}c'_{\tau(s_j)\tau(t)}$.
 Assume w.l.o.g. that $c'_{\tau(s_i),\tau(l)}< c'_{\tau(s_j),\tau(l)}$. Then,
 there exists a $l'\in L\setminus\{l\}$  such that $c'_{\tau(s_i),\tau(l')}>
 c'_{\tau(s_j),\tau(l')}$.
Since $l\neq l'$, then $\tau(l)=\tau'(l)\neq \tau'(l')=\tau(l')$. Assume
w.l.o.g.\ $\tau(l)< \tau(l')$. Then, the Demidenko condition (\ref{dem.s})  for $(C')^{\tau}$
is violated by the indices $\tau(s_i)<\tau(s_j)<\tau(l)< \tau(l')$,
contradicting the selection of $\tau$ as a Demidenko permutation with respect
to $C'$. Analogoulsy the assumption  $\tau(l')< \tau(l)$ leads to a
contradiction to $\tau'$ being a Demidenko permutation for $C'$. 
Thus there exists no entry $l$ in which the columns of $C'$ indexed by $s_i$ and
$s_j$ differ, except may be for the diagonal entries. 
\qed.
\section{A recognition algorithm for  permuted Demidenko matrices}\label{sec:algo}
In this section we present an  $O(n^4)$ algorithm to solve the recognition problem for
Demidenko matrices. The pseudocode  is given  in Algorithm~\ref{RecPermD}
and involves the procedure
 {\tt CheckCandidateDP} presented below.
The basic idea is to exploit  the relationship between Demidenko matrices and
Anti-Robinson matrices described in Section~\ref{sec:relationship} and
to use some algorithm for the recognition of permuted
Anti-Robinson matrices known in the literature, see  \cite{LS,PF} and the
references therein.   More concretely, we use   the
$O(n^2)$   algorithm of Pr\'ea
and Fortin~\cite{PF} for the recognition of  $n\times n$  Anti-Robinson
matrices. In the following  this algorithm is denoted  by {\sc Alg}.
{\sc Alg} takes as  input a symmetric matrix and its size and outputs a logical
variable with the value TRUE,  if the input is  a permuted
Anti-Robinson matrix,  and FALSE otherwise. {\sc Alg} also has  a second output, which
is an
Anti-Robinson permutation for the input matrix, if the logical output is TRUE.
If the logical output is FALSE, the second output of {\sc Alg} is obsolete and 
can be anything. 
Notice that the time complexity $O(n^2)$ of {\sc Alg} is the best possible given that the size of
the input is $O(n^2)$.
\smallskip

\begin{algorithm}
\caption{A recognition algorithm for permuted Demidenko matrices}\label{RecPermD}
\begin{algorithmic}[1]
  \Procedure{RecognPD}{C,n}
\For{$p=1$ to $n$}
\For{$q=1$ to $n$, $p\neq q$}
\State Set $r=0$, $\pi(1)=p$, $\pi(n)=q$.
 \State Set $\pi(i)=0$ for $i\in \{2,3,\ldots,n-1\}$.
 \For{$i=1$ to $n$}
 \For{$i=1$ to $n$}
 \State {Set $c'_{ij}=c_{ij}-c_{pj}-c_{ip}$.}
 \EndFor
 \EndFor
\State (IsDPCandidate,$\pi$)=CheckCandidateDP($C'=(c'_{ij}),n,r,\pi$) \label{CallRecogn}
\If {IsDPCandidate= TRUE}
\If {$(C')^{\pi}$ fulfills conditions (\ref{dem.r})}
\State \label{PD} Return ``$C$ is a permuted Demidenko matrix with a 
Demidenko permutation  $\pi$'' and stop.
\EndIf
\EndIf
\EndFor
\EndFor
\State \label{noPD} Return ``$C$ is not a  permuted Demidenko matrix'' 
\EndProcedure
\end{algorithmic}
\end{algorithm}
\smallskip

 Algorithm~\ref{RecPermD} takes as input  a symmetric $n\times n$
matrix $C$ and its size $n$ and  checks the
existence of a Demidenko permutation $\tau$ for $C$ such that $\tau(1)=p$ and
$\tau(n)=q$ for each fixed pair of indices $p,q\in I:=\{1,2,\ldots, n\}$,
$p\neq q$. This is done  by calling the procedure {\tt CheckCandidateDP} (see line~\ref{CallRecogn}
of Algorithm~\ref{RecPermD}) which has four inputs. The two first inputs are  the  matrix $C'$
(computed as
  specified in Lemma~\ref{lemm:Robinsoncharact1}) and  its size $n$.
  The third input is a  {\sl  partial permutation}  $\pi$.   {\tt CheckCandidateDP} checks  whether
  $\pi$ can be completed to a 
  candidate Demidenko permutation   for $C$. At the first call of {\tt
    CheckCandidateDP} $\pi$ is  initialized  by setting $\pi(1)=p$, $\pi(n)=q$ and $\pi(i)=0$, for
  $i\not\in\{2,\ldots,n-1\}$. Thus, at initialization only two entries of the
  partial permutation  $\pi$
  are fixed. During (recursive) calls of  {\tt CheckCandidateDP}   further entries of $\pi$ are fixed.
  Indeed, the third input of {\tt CheckCandidateDP} is
  the number $r$ of   fixed  entries of $\pi$ besides  $\pi(n)$. {\tt
    CheckCandidateDP}$(C',n,r,\pi)$   returns two values,  a logical
  value IsDPCandidate which is TRUE, if it is possible to complete $\pi$ to  a candidate Demidenko
  permutation for $C$, and  FALSE otherwise. The second output is $\pi$; it 
  is the candidate  Demidenko permutation for matrix $C'$ (and also $C$), if IsDPCandidate is returned
  TRUE, and an (obsolete) partial permutation, if IsDPCandidate is returned FALSE.
  Assume that {\tt CheckCandidateDP} returns (TRUE,$\pi$). In this case Lemmas \ref{lemm:firstorder1} and \ref{lemm:Robinsoncharact1}
   guarantee that $\pi$ is a Demidenko permutation for $C'$ (and thus for $C$)
   provided that there exists a Demidenko permutation for $C$ which maps $1$
   and $n$ to the current values of $p$ and $q$, respectively.
   Thus,    Algorithm~\ref{RecPermD} checks whether $\pi$ is a Demidenko permutation for
   $C'$  and 
   stops with the  corresponding message in the positive case (see line~\ref{PD}  of
   Algorithm~\ref{RecPermD}).
   In the negative case, there is no Demidenko permutation for $C$
   which maps $1$ to $p$ and $n$ to    $q$ (cf.\ Lemmas \ref{lemm:firstorder1} and   \ref{lemm:Robinsoncharact1}).
  Then,  {\tt CheckCandidateDP} is called for
  the next pair $(p,q)$. If IsDPCandidate still equals FALSE at the end of the double for--loop,
 then there is no Demidenko permutation for $C$ and Algorithm~\ref{RecPermD}
  stops with the corresponding message in  line~\ref{noPD}.
  \smallskip

The procedure  {\tt CheckCandidateDP} first
computes the quantities $m:=\min \{ \sum_{x=1}^r c_{i\pi(x)}-rc_{i\pi(n)}\colon i
  \in I'\}$ and $K:=\{ i \in
  I' \colon \sum_{x=1}^r c_{i\pi(x)}-rc_{i\pi(n)}=m\}$
where $I'=\{1,2,\ldots,n\}\setminus \{\pi(i)\colon i \in \{1,2,\ldots,r\}\cup \{n\}\}$
(as specified in
  Lemma~\ref{lemm:firstorder1}). Then,  the following three cases are distinguished:
  $|K|=1$, $|K|=n-r-1$ and $1<|K|<n-r-1$.
  \smallskip
  
  In the case $|K|=1$ $K=\{s\}$, the algorithm  sets
  $\tau(r+1)=s$. In this case,  Lemma~\ref{lemm:firstorder1} implies
  that   $\tau(r+1)=s$ must hold for  every Demidenko permutation $\tau$ 
  for  $C$ mapping the indices $\{1,2,\ldots, r,n\}$
as specified by the incomplete permutation $\pi$. Then $\pi$ and $r$ are updated accordingly and a
  recursive call of  {\tt CheckCandidateDP} follows. 
  \smallskip
  
  In the case  $|K|=n-r-1$,  {\sc
    Alg} is applied  to check whether $C'[K]$ defined as in
  Lemma~\ref{lemm:Robinsoncharact1}(i) is an Anti-Robinson matrix.
    In the negative  case, i.e.\ if {\sc Alg} returns FALSE,  Lemma~\ref{lemm:Robinsoncharact1}(i) implies that there exists  no
  Demidenko permutation for $C$ which maps the indices $\{1,2,\ldots, r,n\}$
as specified by the incomplete permutation $\pi$. Accordingly, {\tt CheckCandidateDP} returns FALSE.
 If {\sc Alg} returns TRUE, then  $\psi\in {\cal S}_{n-r-1}$ is an   Anti-Robinson permutation with respect
  to  $C'[K]$. According to  Lemma~\ref{lemm:Robinsoncharact1}(i) the partial
  permutation $\pi$ can then be completed  to a candidate Demidenko permutation for
  $C$  which coincides with  $\psi$ on $K$. {\tt CheckCandidateDP} returns
  TRUE and the candidate Demidenko permutation $\pi$ for $C$. 
\smallskip
  
  Finally, in the case  $1<|K|<n-r-1$,  with $K=\{s_1,s_2,\ldots,s_{|K|}\}$ and
  $s_1\le s_2\le \ldots \le s_{|K|}$, the $(|K|+1)\times (|K|+1)$ matrix $D$ is constructed
  as specified
in  Lemma~\ref{lemm:Robinsoncharact1}(ii) (lines~\ref{constructD:begin} to
\ref{constructD:end} in procedure {\tt CheckCandidateDP}).  Then,  {\sc Alg} is
applied to check whether
$D$ is a permuted Anti-Robinson matrix.
In the negative case,
Lemma~\ref{lemm:Robinsoncharact1}(ii) implies that there is  no  Demidenko
permutation  for $C$ which maps the indices $\{1,2,\ldots, r,n\}$
as specified by the incomplete permutation $\pi$.  Accordingly, {\tt CheckCandidateDP} returns FALSE. 
In the positive case   {\tt CheckCandidateDP} fixes the values of $\pi$ for
further $|K|$ indices (line~\ref{further:fixPi} of {\tt CheckCandidateDP}) in
accordance with the second statement in Lemma~\ref{lemm:Robinsoncharact1}(ii).
Then $\pi$ and $r$ are updated accordingly and a
  recursive call of  {\tt CheckCandidateDP} follows. 
  \smallskip
  
Summarizing,  we conclude that Algorithm~\ref{RecPermD} correctly decides whether its
input is a permuted Demidenko matrix and   outputs  a Demidenko
permutation for the input matrix  in the positive case. 
\newpage

\begin{algorithmic}[1]
  \label{procedure}
\Procedure{CheckCandidateDP}{$C,n,r,\pi$}
\State Set $I':=\{1,2,\ldots,n\}\setminus \{\pi(i)\colon i \in \{1,2,\ldots,r\}\cup \{n\}\}$.
\State Set $m:=\min \{ \sum_{x=1}^r c_{i\pi(x)}-rc_{i\pi(n)}\colon i
  \in I'\}$.\label{sums:line}
  \State Set $K:=\{ i \in
  I' \colon \sum_{x=1}^r c_{i\pi(x)}-rc_{i\pi(n)}=m\}$.
\If{$|K|=1$}
 \State Set $\pi(r+1)=x$ for $x\in K$.
 \State Set $r:=r+1$.
\State  (IsDPCandidate,$\pi$)=CheckCandidateDP($C,n,r,\pi)$
\EndIf
\If {$|K|=n-r-1$}
\State Compute $C[K]$ for  $K=\{s_1,s_2,\ldots,s_{|K|}\}$.
 \State (IsDPCandidate,$\psi$)={\sc ALG}$(C[K],|K|)$ \Comment{Check whether $C[K]$ is a
   permuted Anti-Robinson matrix}
 \If {IsDPCandidate=TRUE}
 \For{$i=1$ to $n-r-1$}
 \State Set $\pi(r+i)=s_{\psi(i)}$.
 \EndFor
 \EndIf
\State Return (IsDPCandidate,$\pi$)
\EndIf
\If {$1<|K|<n-r-1$}
\State Set $M:=2\sum_{i=1}\sum_{j=1}^n |c_{ij}|$ \Comment{Generate a large
  number $M$}
\For {$i=1$ to $|K|$}\Comment{Construct the 
  matrix $D$  as in Lemma~\ref{lemm:Robinsoncharact1}}\label{constructD:begin}
\For {$j=1$ to $|K|$}
\State Set $d_{ij}:= c_{s_i,s_j}$  
\EndFor
\EndFor
\For {$i=1$ to $|K|$}
\State  $d_{i,|K|+1}:=M\sum_{l\in I'\setminus K}c_{s_il}$,
$d_{|K|+1,i}:=d_{i,|K|+1}$
\EndFor
\State Set $d_{|K|+1, |K|+1}:=M$\label{constructD:end}
\State (IsDPCandidate,$\psi$)={\sc ALG}$(D,|K|+1)$ \Comment{Check whether $D$ is a
   permuted Anti-Robinson matrix}
\If {IsDPCandidate=TRUE}
 \For{$i=1$ to $|K|$}
 \State Set $\pi(r+i)=s_{\psi(i)}$ \label{further:fixPi}
 \EndFor
 \State Set $r:=r+|K|$
 \State  (IsDPCandidate,$\pi$)=CheckCandidateDP($C,n,r,\pi)$
 \EndIf
\State Return (IsDPCandidate,$\pi$)
\EndIf
\EndProcedure
\end{algorithmic}
\newpage

\noindent{\bf The  complexity of the algorithm.}  We observe that
Algorithm~\ref{RecPermD} can be implemented to run in $O(n^2)$ time 
 for each fixed pair of indices $(p,q)$, thus
 implying a total time complexity of $O(n^4)$.
Indeed, consider the computational effort needed for  a fixed pair of
indices $(p,q)$. The computation of $C'$ trivially takes $O(n^2)$ time. 
{\tt CheckCandidateDP}  includes two
major operations: the identification of the index set $K$ 
and the recognition of a permuted
Anti-Robinson matrix ($C[K]$ or $D$).
The identification of $K$ relies on the computation of the sums   in
line~\ref{sums:line} of  {\tt CheckCandidateDP}. Notice that during {\tt
  CheckCandidateDP} new  values of  the partial
permutation $\pi$ are fixed,  but existing values are never changed. Thus, the
  sums mentioned above can be incrementally 
  computed  in a total  of  $O(n^2)$ time  for all recursive calls   of {\tt
  CheckCandidateDP}.
The computation of $m$ and  $K$  can be clearly done in $O(n)$ time for every single
call of  {\tt CheckCandidateDP}.
By observing that   the  recursive calls of   {\tt CheckCandidateDP} 
   operate with pairwise disjoint subsets $K$
  of $\{1,2\ldots,n\}$, we conclude that there are at most $n$ such calls.
  Consequently, the computation of $m$ and $K$  can be done in  a total of
  $O(n^2)$    time. 
Consider finally  the computational effort incurred by all calls of {\sc Alg}.
 Since {\sc Alg} is applied to  pairwise disjoint submatrices of
 $C'$,
 $\sum_{i=1}^ln_i=n$,
 and runs in quadratic time,  we get an
overall time  of   $O(n^2)$.
\smallskip

Summarizing we obtain the following theorem.
\begin{theorem}
Algorithm~\ref{RecPermD} correctly solves the recognition problem for the class
of Demidenko matrices. It can be implemented to run in $O(n^4)$ time where $n$
is the size of the input matrix. 
  \end{theorem}
  
\section{Summary}\label{sec:concl}       
\nopagebreak       
In this paper we have presented  an $O(n^4)$  algorithm for       
the recognition of permuted Demidenko matrices of size $n$, thus resolving a
problem which has been open for  several decades.
This algorithm  closes  a remarkable gap
in the context of recognition problems.
Indeed, the recognition problem has already been   solved for quite a number of
subclasses of the class  Demidenko matrices, e.g.\   the  Kalmanson
matrices, the Supnick matrices, the Monge matrices and the Anti-Robinson
matrices, while remaining open for their common superclass, namely  the  Demidenko
matrices.

Our  algorithm is theoretically   based on the relationship between Demidenko
matrices and Anti-Robinson matrices. It makes use of known algorithms for the
recognition of permuted Anti-Robinson matrices, in particular  \cite{PF}.
The efficient recognition of permuted Demidenko matrices enlarges the class of
polynomially solvable cases of some well known  combinatorial optimization
problems such as the TSP or the Path-TSP. 

\end{document}